\documentclass[aps,prl, superscriptaddress, twocolumn]{revtex4}

\usepackage{bbm}
\usepackage{amssymb}
\usepackage{amsmath}
\usepackage{verbatim}
\usepackage[pdftex]{graphicx}
\usepackage{dsfont}
\usepackage{color}
\usepackage{multirow}

   {\hspace*{\fill}$\Box$\vspace{1.5ex}\par}

\newcommand{\ad}{a^{\dagger}}

\newcommand{\tr}{\mbox{tr}}

\newcommand{\vac}{|0\rangle }

\begin{document}
\title{Generalized Hartree Fock Theory for Dispersion Relations of Interacting Fermionic Lattice Systems}
\author{Christina V.\ Kraus}
\affiliation{Institute for Quantum Optics and Quantum Information of the Austrian Academy of Sciences, A-6020 Innsbruck, Austria}
\affiliation{Institute for Theoretical Physics, University of Innsbruck, A-6020 Innsbruck, Austria}

\author{Tobias J.\ Osborne}
\affiliation{Leibniz Universit\"at Hannover, Institute of Theoretical Physics, Appelstrasse 2, D-30167 Hannover, Germany}

\begin{abstract}
	We study the variational solution of generic interacting fermionic lattice systems using \emph{fermionic Gaussian states} and show that the process of \emph{gaussification}, leading to a nonlinear closed equation of motion for the covariance matrix, is locally optimal in time by relating it to the \emph{time-dependent variational principle}. By linearising our nonlinear equation of motion around the ground-state fixed point we describe a method to study low-lying excited states leading to a variational method to determine the dispersion relations of generic interacting fermionic lattice systems. This procedure is applied to study the attractive and repulsive Hubbard model on a two-dimensional lattice.  
\end{abstract}
\maketitle
The low-lying excited states of a quantum system are of central importance because they are, in contrast to the ground state, accessible in experiments. For example, spectral functions can be directly measured, revealing the particle content of a system. However, many theoretical models involve complicated many-body interactions, and an exact solution for their ground state and excitations are not known. In these cases, we must rely on numerical methods to find good approximations to the low-energy states of the system.

In the case of one-dimensional lattice systems, the Density Matrix Renormalization group (DMRG) \cite{WhiteDMRG, SchollwoeckDMRG} has emerged as the most successful numerical method. By understanding the DMRG as an application of the variational method to the \emph{Matrix Product State} (MPS) \cite{MurgMPS, VerstraeteMPS} class, many generalizations have been recently proposed. This growing collection of powerful algorithms now allows for the determination of the ground state, low-lying eigenstates, and time-evolution of generic one-dimensional quantum spin systems~\cite{Jutho1, Jutho2}.  

Comparatively less progress has been achieved in understanding the low-energy physics of fermionic systems; fermions are the building blocks of all matter, and are thus central to many exciting effects in condensed matter physics, including superconductivity, superfluidity, and the Quantum Hall effect. While an MPS-based approach can be directly applied to one-dimensional fermionic systems, via the Jordan Wigner transformation, this is no longer possible for higher-dimensional systems. Here, fermionic tensor networks, including fermionic Projected Entangled Pair States (PEPS)~\cite{fPEPS} and the Multiscale Entanglement Renormalisation Ansatz (MERA)\cite{MERA} have been recently proposed for their study. However, so far, none of these approaches have been applied to describe the excitations of a fermionic quantum system. 

Another possibility to find ground-state approximations of fermionic systems is to use other variational trial wave functions such as Slater determinants (Hartree Fock theory) or Gutzwiller projected wave functions. These have also been used to describe excitations in the past, in the framework of time-dependent Hartree Fock theory (see e.g. \cite{Dirac, HFT_excitations}) and time-dependent Gutzwiller theory  \cite{Seibold}.  Recently, a novel technique based on the variational class of fermionic Gaussian states (fGS)\cite{LinearOptics} has been proposed \cite{gHFTnum}. This method allows one to find approximate solutions for ground- and thermal states, as well as the time evolution of interacting fermionic lattice systems in any dimension and geometry. An application to the $2d$ spinfull Fermi Hubbard mode has established that these algorithms are stable and efficient, and provided results in agreement with, and going beyond, Quantum Monte Carlo. The fGS class generalizes the basic building blocks of our understanding of fermionic matter, namely BCS-states for superfluid phases, and Slater determinants (Hartree Fock Theory) for Mott and fermionic spin states to the framework of generalized Hartree Fock Theory (gHFT)\cite{bach-1993}, which is a strict superset of Hartree-Fock theory.  As fGS can be described using a number of variational parameters scaling polynomially in the system size, it allows the efficient simulation of large systems.

In this Letter, inspired by the utility of the fGS class for capturing a wide range of physically relevant phases, we take the next step and attempt to describe the low-lying excited states of interacting fermionic lattice systems. Hence we begin by introducing the general model of interacting fermions we investigate in the framework of gHFT. Then we apply the Time Dependent Variational Principle (TDVP) \cite{Dirac, TDVP} to derive a locally (in time) optimal effective equation for the time evolution within the fGS variational class. An expansion of this equation around the ground-state solution thus leads to an approximation of the low-lying excited states. We demonstrate the power of our approach by investigating the two-dimensional Hubbard model, both in the attractive and repulsive regimes, deriving the excitation spectrum, and discussing the nature of the excitations. 

We consider a system of $M$ fermionic modes localised on a lattice in $d$ dimensions. This is described by fermionic creation and annihilation operators obeying the canonical anti-commutation relations (CAR), $\{\ad_k, a_l \} = \delta_{k,l}$, $\{a_k, a_l\} =0$. Such a system can be equivalently described by $2M$ real \emph{Majorana operators}, $c_j = \ad_j + a_j$ and $c_{j+M} = (-i)(\ad_j - a_j)$ obeying the CAR $\{c_j, c_k\}= 2 \delta_{jk}$. Many systems of importance in condensed-matter physics are modeled by Hamiltonians involving two-body interactions. In the Majorana language, the most general such model is written as 
\begin{align}\label{eq:intHam}
H = i\sum_{kl}T_{kl}c_kc_l + \sum_{klmn}U_{klmn}c_kc_lc_mc_n,
\end{align}
where $T = -T^T \in \mathds{R}^{2M\time 2M}$ and $U_{klmn} \in \mathds{R}$ is antisymmetric under the exchange of any two adjacent indices. 

Recently, in \cite{gHFTnum}, the fGS variational class has been used to find approximate solutions for the ground- and thermal states of \eqref{eq:intHam}. The fGS class is defined as the set of all states which are exponentials of a quadratic form in the fermionic operators. Such states fulfill Wick's theorem, and thus can be completely described on a single-particle level via the real and antisymmetric covariance matrix (CM) $\Gamma_{kl} = i/2\langle [c_k, c_l]\rangle$. For physical states the CM obeys the inequality $i \Gamma \leq \mathds{1}$, while for pure states we have $\Gamma^2 = - \mathds{1}$. Note that every pure fGS is the ground state of a quadratic Hamiltonian $H = i \sum_{kl}h_{kl}c_kc_l$, where $h = -h^T \in \mathds{R}$. Further, Gaussian states remain Gaussian under time evolution according to any quadratic Hamiltonian, so that the time evolution can be formulated in terms of the CM alone as $\dot \Gamma(t) = 4[h, \Gamma(t)]$.

We briefly review how to approximate the time evolution of an interacting system of the form \eqref{eq:intHam} using fGS. Since $H$ generically includes nonquadratic interactions, any infinitesimal time step $\Delta t$ takes us out of the fGS class. Thus, we have to \emph{project back} into the set of fGS after each time step. In \cite{gHFTnum}, this was done via the process of \emph{gaussification}, where Wick's theorem is invoked after each time step to reexpress the system's state as an fGS. In this way, a closed evolution of the CM could be derived: $\dot \Gamma(t) = 4[h^{(6)}(\Gamma(t)), \Gamma(t)]$, where $h^{(6)}(\Gamma) = T + 6\tr_2[U\Gamma]$. Thus time evolution can be formulated as the \emph{nonlinear} evolution according to a quadratic but \emph{state-dependent} Hamiltonian. In the following, we use the \emph{Time-dependent Variational Principle} to prove that gaussification, used \emph{ad hoc} in \cite{gHFTnum}, is actually \emph{locally optimal in time} within the set of fGS.  

\begin{figure}[tb]
\includegraphics[width=0.6\columnwidth]{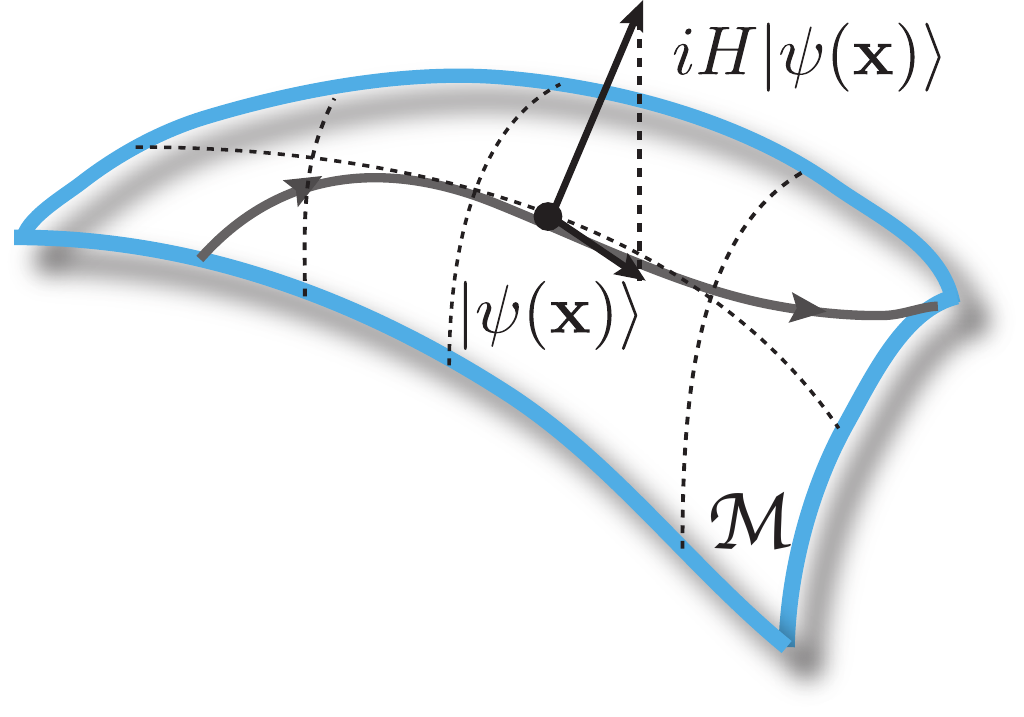}
\caption{Graphical representation of the TDVP within the variational manifold $\mathcal{M}$ of Gaussian states represented by state vectors $|\psi(\mathbf{x})\rangle$. The time evolved vector $iH|\psi(\mathbf{x})\rangle$ is in general not an element of the manifold $\mathcal{M}$, and we have to find an optimal approximation within the variational manifold. Thus, we obtain the optimal path $|\psi(\mathbf{x})\rangle$ within the manifold $\mathcal{M}$.\label{fig:TDVP}}
\end{figure}

We briefly recall the TDVP. Suppose we have some variational class of states $\{\vert \psi(\mathbf{x})\rangle\,\vert\, \mathbf{x}\in \mathds{R}^D\}$. We aim to solve the time-dependent Schr{\"o}dinger equation $d/dt |\psi(\mathbf{x})\rangle = -iH |\psi(\mathbf{x})\rangle$ as best as possible while remaining in our class. Unfortunately, in general, the vector $iH |\psi(\mathbf{x})\rangle$ is not an element of the tangent space, and we can only find an approximate solution. The optimal way to do this is to carry out the minimisation $\min_{\dot{\mathbf{x}}}|| \dot x^j \partial_j |\psi(\mathbf{x})\rangle + iH |\psi(\mathbf{x})\rangle||$, were we have written $|\psi(\mathbf{x})\rangle$ as a linear combination of vectors in the tangent space, and $\partial_j = \partial/\partial x_j$. If we perform this optimisation at each time step, and take the limit of infinitesimal step size, we arrive at the \emph{locally optimal} evolution within our variational class. Note that this minimization is equivalent to solving for the Euler-Lagrange Equation $\tfrac{d}{dt} \tfrac{\partial \mathcal{L}}{\partial \dot x_j} - \tfrac{\partial \mathcal{L}}{\partial  x_j} = 0$ with Lagrangian $\mathcal{L} = \langle \psi(\mathbf{x})|\left(-i \tfrac{d}{dt} - H \right)|\psi(\mathbf{x})\rangle$. 

We now show that gaussification is locally optimal by demonstrating its equivalence to the TDVP. To see this, note that any pure fermionic Gaussian state can be represented as $|\psi_G(t)\rangle = \mathcal T \exp \left[\frac{i}{2}\int_0^t \sum_{k,l}G_{kl}(t') c_k c_l\right] \vac$, where $G = -G^T$ is a real matrix. Thus $\Gamma$ and $G$ are related via $\dot \Gamma = [G, \Gamma]$. This relation is imposed in a Lagrangian using a Lagrange multiplier $\Lambda$, and we obtain for the TDVP Lagrangian the expression $\mathcal{L} = \langle \psi(\mathbf{x})|\left(-i \tfrac{d}{dt} - H \right)|\psi(\mathbf{x})\rangle = \tr[\Gamma(G + h^{(3)}(\Gamma))] + \tr[\Lambda(\dot \Gamma - [G, \Gamma])]$, where $h^{(3)}(\Gamma) = T + 3\tr_2[U\Gamma]$. The Euler-Lagrange equations of motion immediately lead to $\dot \Gamma = 4[h^{(6)}, \Gamma]$, which are exactly those arising from gaussification in Ref. \cite{gHFTnum}. Thus we obtain a locally optimal smooth path in our variational manifold, as depicted in Fig.~\ref{fig:TDVP}.

The optimality of the TDVP is now exploited to derive an approximation of the excitation spectrum within the fGS variational class. To this end we linearise our TDVP equation of motion around the variational ground state given by $\Gamma_0$, i.e., we write $\Gamma(t) = \Gamma_0 + \epsilon \Gamma_1 + \mathcal{O}(\epsilon^2)$, with real $\Gamma_1 = -\Gamma_1^T$. [Note that $\Gamma_0$ can be obtained via imaginary time evolution with respect to the quadratic, but state dependent, Hamiltonian $H = i \sum_{k,l}h^{(6)}c_kc_l$ \cite{gHFTnum}: starting from an arbitrary pure state $\Gamma(0)$ the evolution $\Gamma(t) = O(t) \Gamma(0)O(t)^T$ where $O(t) = \mathcal{T}\exp\left[\int_0^t dt^{\prime}2 [\Gamma(t'), h^{(6)}(\Gamma(t'))]\right] $ leads to $\Gamma_0$.] We can eliminate the time dependence by writing $\Gamma_1(t) = e^{i\omega t}\hat\Gamma(\omega)$ and arrive at the following eigenvalue equation, which is the main result of this work:
\begin{align}\label{eq:excitation}
i\omega\hat \Gamma_1(\omega) = [h^{(6)}(\Gamma_0), \hat \Gamma_1(\omega)] + 6 [\tr[U \hat \Gamma_1(\omega)], \Gamma_0].
\end{align}
While  \eqref{eq:excitation} gives the general solution for the excitation spectrum within the fGS variational class, this matrix equation is, in general, hard to solve.  The problem may be simplified by rewriting it as an eigenvalue equation via \emph{vectorization}:
\begin{align}
i\omega |\Gamma_1 \rangle &= \mathcal{V} |\Gamma_1\rangle,\nonumber\\
\mathcal{V} &= h^{(6)}\otimes \mathds{1} + \mathds{1}\otimes h^{(6)} - 6(\Gamma\otimes \mathds{1} + \mathds{1}\otimes \Gamma)U,
\end{align}
where $\mathcal{V}$ is a $4M^2 \times 4M^2$ matrix. Thus, the problem reduces to diagonalizing the matrix $\mathcal{V}$, which can be numerically demanding for large systems. The problem is easier in the presence of additional symmetries, however: in the case of a translation invariant system the problem simplifies considerably. In this case the matrix $\Gamma_1$ is block-diagonal in Fourier space, i.e.\ $\hat \Gamma_1 = \bigoplus_{\mathbf{k}} \Gamma_1(\mathbf{k})$, with modes $\mathbf{k}$ and $-\mathbf{k}$ paired. Thus, for each momentum mode $\mathbf{k}$, we only need to solve the reduced equation $i \omega(\mathbf{k})|\Gamma_1(\mathbf{k})\rangle = \mathcal{V}(\mathbf{k})|\Gamma_1(\mathbf{k})\rangle$. In this way we can extract the \emph{dispersion relation} $\omega(\mathbf{k})$. Further, since the CM contains all information about a fGS, we can also infer the nature of the excitations by looking at $\Gamma_1(\mathbf{k})$. 

For the remainder of this Letter we apply our approach to the two-dimensional Hubbard model on a square lattice. The Hubbard model describes an interacting fermionic lattice system of particles with two internal spin states, $ \sigma = \uparrow, \downarrow$,
\begin{align}
H_{\text{Hubb}} = -t\sum_{\langle \mathbf{x,y}\rangle, \sigma} \ad_{\mathbf{x},\sigma} a_{\mathbf{y},\sigma}  + u\sum_{\mathbf{x}}n_{\mathbf{x} \uparrow}n_{\mathbf{x} \downarrow} - \mu \sum_{\mathbf{x}, \sigma}n_{\mathbf{x} \sigma},
\end{align} 
where $\mathbf{x}$ denotes a position on the lattice, $\langle ...\rangle$ indicates a summation over nearest neighbors, and $n_{\mathbf{x} \sigma} = \ad_{\mathbf{x} \sigma}a_{\mathbf{x} \sigma}$ is a particle number operator. We take the hopping parameter $t$ to be real and consider $u$ positive (negative) in case of a repulsive (attractive) interaction. The chemical potential $\mu$ fixes the filling of the lattice. 

Despite its simple structure the Hubbard model allows for a wide range of physical phases, including Mott and spin ordered phases in the case of a repulsive interaction, and superfluid phases in case of an attractive interaction. It is even believed that the doped Hubbard model at positive $u$ may provide a description of high-temperature superconductivity. However, unless we consider very special parameter regimes, an exact solution of the model is unknown, and despite an intense theoretical and numerical effort the precise structure of its phase diagram remains an open question.  

In the following we determine the excitation spectrum of the translation-invariant Hubbard model with periodic boundary conditions (PBC) in two dimensions within the fGS variational class using Eq.~\eqref{eq:excitation}. To calculate $\mathcal{V}$ we need the CM of the ground state, $\Gamma_0$. For the translation invariant case with PBC it has been shown in \cite{bach-1993} that $\Gamma_0$ can be obtained via a two-parameter optimization for aribitrary filling, both in the attractive and repulsive regimes. We use this result to numerically determine $\Gamma_0$ on a $31\times 31$ lattice, transform into Fourier space, and are left, for each $\mathbf{k}$, with the diagonalization of an $8 \times 8$ matrix. 

Let's discuss our results for the Hubbard model in the attractive case for three exemplary sets of parameters $(u, \mu) = (-4, 1), (-2, 2), (-4, 3)$. In Table~\ref{table:GSproperties_uneg} we have summarised various properties of the ground states. We have calculated the filling $n = N/(2\cdot 31^2)$ and the pairing per particle~\cite{Pairing}, $p = \sum_{i,j, \sigma, \sigma'}|\langle \ad_{i\sigma}\ad_{j\sigma'}\rangle|^2/N$, where $N$ is the number of particles in the lattice. 

 \begin{table}[h]
\begin{tabular}{| c| c | c | c |}
\hline
$(u, \mu)$ & $ (-4, 1)$ &  $ (-2, 2)$ &  $ (-4, 3)$ \\\hline
$n$           &  0.83 & 0.24 & 0.17 \\\hline 
$p$           & 0.028 & 0.045 & 0.13\\ \hline
\end{tabular} 
\caption{Filling $n$ and pairing per particle $p$ for the attractive Hubbard model on a $31 \times 31$ lattice.\label{table:GSproperties_uneg}}
\end{table} 
We consider configurations far from half-filling. This is because at half filling the Hubbard model has additional symmetries leading to a highly degenerate ground state which severely complicates analysis of the excitations. The attractive Hubbard model supports superfluid phases indicated by a non-vanishing pairing per particle. Thus, the ground state is gapped for all three sets of parameters. 

\begin{figure}[tbh]
\includegraphics[width=0.9\columnwidth]{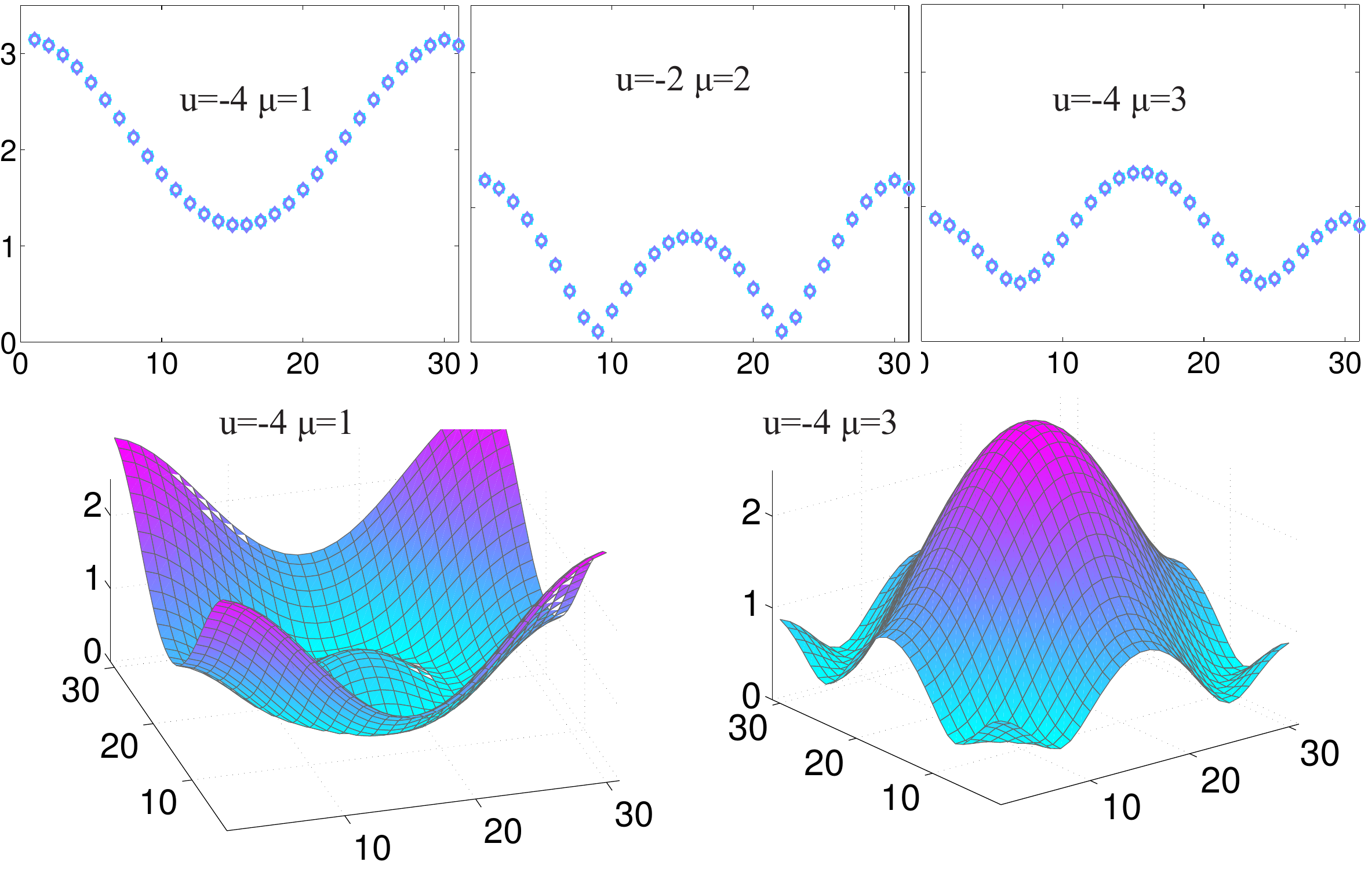}
\caption{(Color online) Dispersion relation for the negative u Hubbard model \label{fig:Hubb_uneg}}%
\end{figure}

Next, we investigate the excitation spectra. The results are depicted in Fig.~\ref{fig:Hubb_uneg}.  We see that for $(u,\mu) = (-4,1)$ and $(u,\mu) = (-4, 3)$ the excitation spectrum is gapped, while it is gapless for  $(u,\mu) = (-2, 2)$. Even though the results shown in Fig.~\ref{fig:Hubb_uneg} seem to imply that there is only one possible excitation, a closer investigation reveals that for all three sets of parameters we have six non-trivial dispersion relations $\omega_j(\mathbf k)$, $j = 1, \ldots, 6$, with $\omega_1(\mathbf{k})\leq \omega_2(\mathbf{k})\leq \ldots$. We find for all three sets of parameters that $\omega_2 = \omega_3$ and $\omega_4 = \omega_5 = \omega_6$.  In Table~\ref{table:Excitations_uneg} we characterize these excitations via charge, spin and pairing order parameters $C$, $S$ and $\Delta$ respectively.
 \begin{table}[h]
\begin{tabular}{|c | c |c | c | c | c | c|}
\hline
$(u, \mu)$&$\Delta_{k=0}$&$ \Delta_k$ & $\Delta_S$& $S_{z}$&$S_T$  & $C$\\\hline
$(-4, 1), (-4, 4)$ &$\omega_1$, $\omega_{4-6}$ & $\omega_{2-3}$ & $\omega_{4-6}$ &  $\omega_{4-6} $   &  $\omega_{4-6} $  & $\omega_1$\\\hline
$(-2, 2)$ &$\omega_1$, $\omega_{4-5}^{*}$ & $\omega_{2-3}$ & $\omega_{4-5}$ &  $\omega_{4-5}^* $   &  $\omega_{4-6} $  & $\omega_1$\\\hline
 \multicolumn{7}{|c|} {$\Delta_{k=0} = \langle \ad_{k\uparrow} \ad_{-k \downarrow}\rangle $, $\Delta_{k} = \langle \ad_{k\uparrow} \ad_{k \downarrow}\rangle$, $\Delta_{S} = \langle \ad_{k\uparrow} \ad_{-k \uparrow}\rangle$}\\
 \multicolumn{7}{|c|} {$S_T = \langle \ad_{k \uparrow} a_{-k\downarrow}\rangle$,$S_z = \langle n_{k \uparrow} - n_{k\downarrow}\rangle$, $C = \langle n_{k\uparrow} + n_{k\downarrow}\rangle$}\\\hline
\end{tabular} 
\caption{Classification of the excitations of the attractive Hubbard model. We distinguish charge ($C$), spin ($S$) and pairing ($\Delta$) excitations and order $\omega_1(\mathbf{k})\leq \omega_2(\mathbf{k})\leq \ldots$. We find 6 non-trivial excitations in all three cases. Excitations $j_1 < \ldots < j_2$ with the same dispersion are labeled by $\omega_{j_1 - j_2}$. $\omega_{4-5}^*$: For $(u, \mu) = (-2, 2)$ we find that $\omega_6$ has a spin and pairing order parameter of less than $1\%$ compared to $\omega_{4-5}$ and is thus set to zero in the table.\label{table:Excitations_uneg}}
\end{table} 
We find that for $(u,\mu) = (-4, 1), (-4, 3)$ the excitations are always gapped, while they become gapless for $(u, \mu) = (-2, 2)$. Here, $\omega_6$ has a pairing $\Delta_{k=0}$ and a spin order $S_z$ of less than $1\%$ compared to the values for $\omega_{4-5}$. 

Next, we discuss excitations in the repulsive Hubbard model. Here, as predicted in~\cite{bach-1993}, the ground state is never paired, i.e. $p = 0$. Spin-ordered phases within a gHFT treatment are only predicted for half-filling, or large $u$. Thus, for our sets of parameters, $(u, \mu) = (4,0), (4, -3), (4, -4)$ we only present the filling $n$ in Table~\ref{table:GSproperties_upos}.

\begin{figure}[t]
\includegraphics[width=0.9\columnwidth]{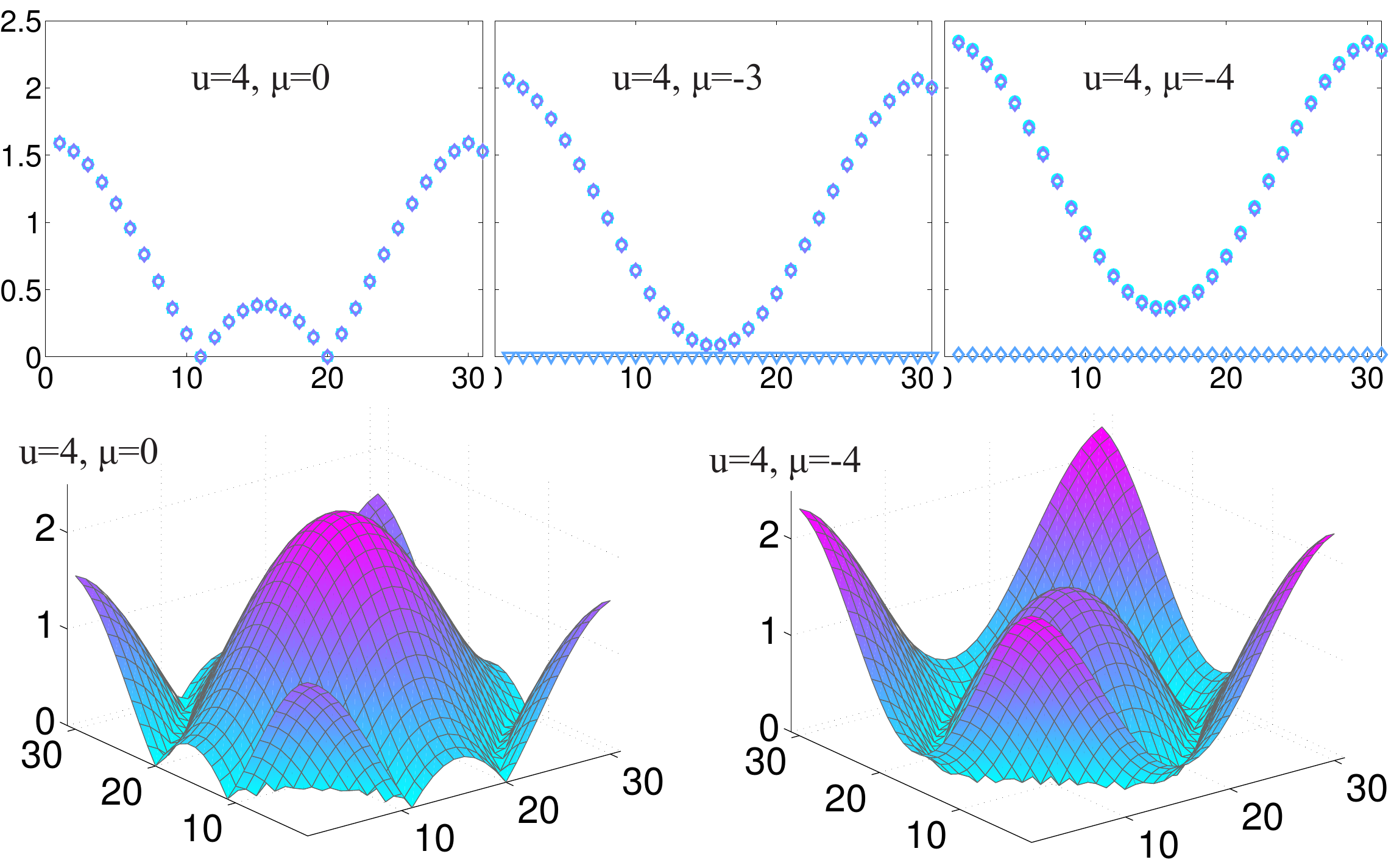}
\caption{(Color online) Dispersion relation for the positive u Hubbard model}%
\label{fig:Hubb_upos}%
\end{figure}

 \begin{table}[h]
\begin{tabular}{| c| c | c | c |}
\hline
$(u, \mu)$ & $ (4, 0)$ &  $ (4, -3)$ &  $ (4, -4)$ \\\hline
$n$           &  0.18 & 0.70 & 0.81 \\\hline 
\end{tabular} 
\caption{Filling $n$ for the repulsive Hubbard model on a $31 \times 31$ lattice.\label{table:GSproperties_upos}}
\end{table} 
Now we discuss the excitation spectrum depicted in Fig.~\ref{fig:Hubb_upos}. We find ten non-trivial excitations for $(u, \mu) = (4, -3), (4, -4)$ [$\omega_{7-10}$ have a flat dispersion], and six for $(u, \mu) = (4,0)$. Again, for each set of parameters there are excitations with different dispersion relations; we summarize the nature of the excitations in Table~\ref{table:Excitations_upos}.

 \begin{table}[h]
\begin{tabular}{|c | c |c | c | c | }
\hline
$(u, \mu)$&$\Delta_{k=0}$&$ \Delta_k$ & $\Delta_S$ &$S_T$ \\\hline
\multirow{2}{*}{$(4, -3), (4, -4)$} &$\omega_1$, $\omega_{2}$,  $\omega_{5}$ & $\omega_{3-4}$&$\omega_1$, $\omega_{2}$& $\omega_1$, $\omega_{2}$\\
 &  $\omega_{6}, \omega_{7-8}$  & $\omega_{9-10}$ & $ \omega_{6}, \omega_{7-8}$ & $ \omega_{6}, \omega_{7-8}$\\\hline
 $(4, 0)$ & $\omega_{1-3}$, $\omega_6$  & $\omega_{4-5}$ & $\omega_{1-3}$  & $\omega_{1-3}$ \\\hline
 \multicolumn{5}{|c|} {$\Delta_{k=0} = \langle \ad_{k\uparrow} \ad_{-k \downarrow}\rangle $, $\Delta_{k} = \langle \ad_{k\uparrow} \ad_{k \downarrow}\rangle$ }\\
 \multicolumn{5}{|c|} {$\Delta_{S} = \langle \ad_{k\uparrow} \ad_{-k \uparrow}\rangle$, $S_T = \langle \ad_{k \uparrow} a_{-k\downarrow}\rangle$}\\\hline
\end{tabular} 
\caption{Classification of the excitations of the attractive Hubbard model. We distinguish spin $(S)$ and pair $(\Delta)$ excitations and order $\omega_1(\mathbf{k})\leq \omega_2(\mathbf{k})\leq \ldots$.. Excitations $j_1 < \ldots < j_2$ with the same dispersion are labeled by $\omega_{j_1 - j_2}$.  $\omega_{7-10}$ are the flat excitations near zero in Fig.~\ref{fig:Hubb_upos}.\label{table:Excitations_upos} }
\end{table} 
In contrast to the attractive Hubbard model, we find no charge excitations and only non-vanishing spin order $S_T$. The nature of the excitations for $(u, \mu) = (4, -3), (4, -4)$ is identical, while we find a different structure for $(u, \mu) = (4, 0)$. Here, the excitations with the highest energy is threefold degenerate, and we find no flat excitations near zero energy. Note also, that for large negative chemical potential the excitations become gapped. 

In summary, we have presented how the the TDVP applied to the set of fermionic Gaussian states allows for a variational approach to determine dispersion relations and the nature of the excitations in interacting fermionic systems. For translational invariant systems, the arising equations scale linearly in the system size, and can thus be applied to large systems in more than one dimension, as we have demonstrated for the example of the $2d$ Hubbard model in both the attractive and the repulsive regimes on a $31 \times 31$ lattice.

\emph{Acknowledgments.---}% 
We thank I. Cirac, J. Haegeman and F. Verstraete for useful discussion. This work was supported, in part, by the cluster of excellence EXC 201 “Quantum Engineering and Space-Time Research”, by the Deutsche Forschungsgemeinschaft (DFG),  the EU grant QFTCMPS and by the Austrian Ministry of Science BMWF as part of the UniInfrastrukturprogramm of the Research Platform Scientific Computing at the University of Innsbruck.

\end{document}